\documentclass{iaus}
\usepackage{graphicx}
\usepackage{times}

\title[bHROS observations of line profiles] 
{bHROS high spectral resolution observations of PN forbidden and 
recombination line profiles}

\author[Barlow et al.]   
{M. J. Barlow$^1$, A. S. Hales$^1$, P. J. Storey$^1$, X.-W. Liu$^2$, Y. G. Tsamis$^1$, M. E. Aderin$^1$}

\affiliation{$^1$Dept. of Physics and Astronomy, University College 
London; $^2$Dept. of Astronomy, Peking University}

\pubyear{2006}
\volume{234}  
\pagerange{119--126}
\date{?? and in revised form ??}
\jname{Planetary Nebulae in our Galaxy and Beyond}
\editors{M. J. Barlow \& R. H. M$\acute{e}$ndez, eds.}
\begin{document}

\maketitle

\begin{abstract}
We have acquired high spectral resolution observations (R=150,000) of the
planetary nebulae NGC~7009 and NGC~6153, using bHROS on Gemini South. 
Observations of this type may provide a key to
understanding why optical recombination lines (ORLs) yield systematically
higher heavy element abundances for photoionized nebulae than do the
classical forbidden collisionally excited lines (CELs) emitted by the same
ions; NGC~7009 and NGC~6153 have notably high ORL/CEL abundance
discrepancy factors (ADFs) of 5 and 10, respectively. Due to the opposite
temperature dependences of ORLs and CELs, ORLs should be
preferentially emitted by colder plasma. Our bHROS observations of 
NGC~7009 reveal that the [O~{\sc iii}] 4363~\AA\ CEL has a FWHM linewidth that
is 1.5 times larger than that shown by O~{\sc ii} ORLs in the same
spectrum, despite the fact that all of these lines are emitted by the
O$^{2+}$ ion.  The bHROS spectra of NGC~6153 also show that its O~{\sc ii}
ORLs have significantly narrower linewidths than do the [O~{\sc iii}]
4363~\AA\ and 5007~\AA\ lines but, in addition, the [O~{\sc iii}]
4363~\AA\ and 5007~\AA\ lines show very different velocity profiles,
implying the presence of large temperature variations in the nebula.


\end{abstract}

\section{Introduction}

H~{\sc ii} region and planetary nebula abundances derived from heavy
element optical recombination lines (ORLs) should in principle be more
accurate than those derived from collisionally excited (forbidden) lines
(CELs) from the same ions, due to the exponential dependence upon the
adopted nebular temperature of CEL abundances. However, a number of
abundance surveys of PNe that we have recently published (Tsamis et al.
2004,  Liu et al. 2004, Wesson et al. 2005) show that ORLs yield
systematically higher heavy element abundances than those obtained from
classical CEL forbidden line analyses, with abundance discrepancy factors
(ADFs) of 1.6-3.0 obtained for most nebulae. For about 5-10\% of nebulae,
the ADFs are so large (5-80) that nebular temperature fluctuations (e.g.
Peimbert 1967) cannot be invoked to reconcile the differences. It is
important to resolve this problem, since standard temperature fluctuation
analysis methods effectively adopt the higher ORL abundances, with
profound implications for the mean abundances that are derived for our own
and other galaxies from H~{\sc ii} regions and PNe.

NGC~7009 and NGC~6153 are nebulae for which we have previously found very
large enhancements in the recombination line abundances of the second-row
elements carbon, nitrogen, oxygen and neon (Liu et al. 1995, Liu et al.
2000), with ORL/CEL ADFs of 5 and 10, respectively. Nebular temperature
fluctuations are unable to account for these effects. However, models that
invoke low-temperature hydrogen-deficient clumps appear viable as an
explanation for the high recombination line abundances that are observed
(see Liu et al. 2000 for more details).  If the heavy element ORLs
originate from much cooler material than do the classical forbidden lines,
then an obvious prediction is that ORLs should have narrower lines widths
than forbidden lines from the same ion. In order to test this prediction,
we have obtained high spectral resolution (R=150,000) observations of
these two high-ADF nebulae using the recently commissioned bHROS echelle
spectrograph on the 8-m Gemini South telescope.

\section{Results}

\begin{figure}
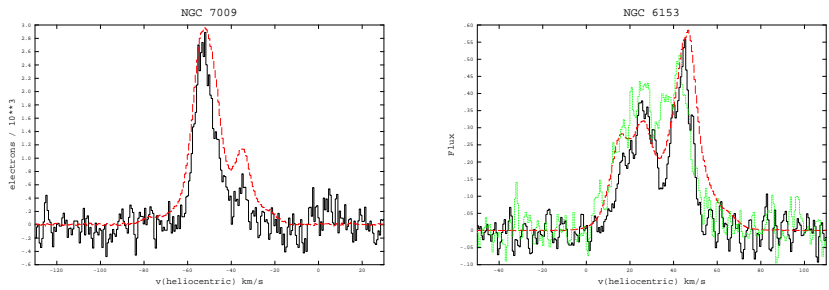

\begin{center}
\includegraphics[angle=-90,width=5.8cm]{barlowf1.eps}
\includegraphics[angle=-90,width=5.8cm]{barlowf2.eps}
\caption{bHROS O{~\sc ii} and [O~{\sc iii}]
profiles for NGC~7009 (left) and NGC~6153 (right). See text for details.}
\end{center}
\end{figure}

Four 30-minute exposures on NGC~7009 were obtained with bHROS on July 27th
2005, placing its 0.9$''\times$~0.9$''$ image-slicer on a bright edge
region located 5.6$''$ NW of the central star, with the echellogram
centred at 4267~\AA .  Fig.~1(a) shows a velocity plot of the co-added
profile of the 4089, 4275 and 4349~\AA\ O~{\sc ii} lines (solid plot),
together with the profile (after division by a factor of ten) of the
[O~{\sc iii}] 4363~\AA\ CEL (dashed). These lines all arise from the
O$^{2+}$ ion, yet the linewidth of the O~{\sc ii} main velocity component
is measured to be 1.47 times smaller than that found for the same velocity
component in the [O~{\sc iii}] line, which for pure velocity broadening
would yield a T$_e$ from the [O~{\sc iii}] line that is 2.2 times larger
than that obtained from the O~{\sc ii} lines. Clearly, the O$^{2+}$ ORLs
and CELs cannot originate from identical material.

Five 30-minute exposures were obtained on NGC~6153 on March 9th 2006 (a
further five exposures were obtained on March 24th; the results shown here
are from the March 9th observations). The 0.9$''\times$~0.9$''$ large bHROS
image-slicer was placed on a bright region located 7.4$''$ SE of the
central star. The echellogram was centred at a longer wavelength (4650~\AA
) than for NGC~7009, to allow for the much higher reddening towards
NGC~6153. Fig.~1(b) shows a velocity plot that compares the observed line
profiles of the O~{\sc ii} 4649~\AA\ ORL (solid), together with those of
the [O~{\sc iii}] 5007~\AA\ (dashed) and 4363~\AA\ (dotted) CELs.  The
5007~\AA\ profile has been divided by a factor of 250 for comparison
purposes. Once again, the O~{\sc ii} ORL line profile shows much narrower
velocity components than do the [O~{\sc iii}] 5007~\AA\ and 4363~\AA\
lines, indicating that the O$^{2+}$ ions giving the O~{\sc ii}
recombination emission are located in physically distinct regions from
those giving the [O~{\sc iii}] forbidden line emission. A big surprise,
though, is that the two [O~{\sc iii}] forbidden lines have very different
velocity profiles. Of the three observed [O~{\sc iii}] 5007~\AA\ velocity
components, the two blue-most components both have [O~{\sc iii}] 4363~\AA\
counterparts, but the gap between the middle and red-most 5007~\AA\
component is infilled by emission in the case of the 4363~\AA\ line, while
only the blue side of the strongest 5007~\AA\ velocity component has a
counterpart in either the [O~{\sc iii}] 4363~\AA\ or O~{\sc ii} 4649~\AA\
profiles. Since the ratio of the [O~{\sc iii}] 5007~\AA\ to 4363~\AA\ line
intensities is a classical electron temperature diagnostic, the large
variations in this ratio as a function of velocity indicate very large
temperature variations within the nebula.


\begin{thebibliography}{}
\bibitem[Liu et al. (1995)]{liu95}
     {Liu, X.-W., Storey, P.J., Barlow, M.J., Clegg, R.E.S.} 1995, \textit{MNRAS} 272, 369
\bibitem[Liu et al. (2000)]{liu00}
     {Liu, X.-W., Storey, P. J., Barlow, M. J., Danziger, I. J., Cohen, M. Bryce, M.} 2000, \textit{MNRAS} 312, 585
\bibitem[Liu et al. (2004)]{liu04}
     {Liu, Y., Liu, X.-W., Barlow, M. J., Luo, S.-G.} 2004, \textit{MNRAS} 353, 1051
\bibitem[Peimbert (1967)]{peim67}	
     {Peimbert, M.} 1967, \textit{ApJ} 150, 825
\bibitem[Tsamis et al. (2004)]{tsam04}
     {Tsamis, Y. G., Barlow, M. J., Liu, X.-W., Storey, P. J., Danziger, I. J.} 2004, \textit{MNRAS} 353, 953
\bibitem[Wesson et al. (2005)]{wess05}
     {Wesson, R., Liu, X.-W., Barlow, M. J.} 2005, \textit{MNRAS} 362, 424
\end{thebibliography}
\end{document}